\documentclass[psf,epsf,superscriptaddress,twocolumn,showpacs,preprintnumbers,floatfix,prb]{revtex4-1}
\usepackage{graphicx}
\usepackage{dcolumn} 
\usepackage{bm}
\usepackage{epsfig}
\begin{document} 

\title{Sr$_6$Co$_5$0{$_1$$_5$}: non-one-dimensional behavior of a charge ordered structurally quasi-one-dimensional oxide.}

\author{A.~S. Botana}
\email[Corresponding author. Email address: ]{antia.sanchez@usc.es}
\affiliation{Departamento de F\'{i}sica Aplicada,
  Universidade de Santiago de Compostela, E-15782 Campus Sur s/n,
  Santiago de Compostela, Spain}
\affiliation{Instituto de Investigaci\'{o}ns Tecnol\'{o}xicas,
  Universidade de Santiago de Compostela, E-15782 Campus Sur s/n,
  Santiago de Compostela, Spain}
\author{P.~M. Botta}
\affiliation{Instituto de Investigaciones en Ciencia y Tecnolog\'ia de Materiales (INTEMA), CONICET-UNMdP, J.B. Justo 4320 B7608FDQ, Mar del Plata, Argentina.}
\author{C. de la Calle} 
\affiliation{Instituto de Ciencia de Materiales de Madrid (ICMM), CSIC, Cantoblanco, 28049 Madrid, Spain.}
\author{A. Pi\~neiro}
\affiliation{Departamento de F\'{i}sica Aplicada,
  Universidade de Santiago de Compostela, E-15782 Campus Sur s/n,
  Santiago de Compostela, Spain}
\affiliation{Instituto de Investigaci\'{o}ns Tecnol\'{o}xicas,
  Universidade de Santiago de Compostela, E-15782 Campus Sur s/n,
  Santiago de Compostela, Spain}
\author{V. Pardo}
\affiliation{Departamento de F\'{i}sica Aplicada,
  Universidade de Santiago de Compostela, E-15782 Campus Sur s/n,
  Santiago de Compostela, Spain}
\affiliation{Instituto de Investigaci\'{o}ns Tecnol\'{o}xicas,
  Universidade de Santiago de Compostela, E-15782 Campus Sur s/n,
  Santiago de Compostela, Spain}
\author{D. Baldomir}
\affiliation{Departamento de F\'{i}sica Aplicada,
  Universidade de Santiago de Compostela, E-15782 Campus Sur s/n,
  Santiago de Compostela, Spain}
\affiliation{Instituto de Investigaci\'{o}ns Tecnol\'{o}xicas,
  Universidade de Santiago de Compostela, E-15782 Campus Sur s/n,
  Santiago de Compostela, Spain}
\author{J.~A. Alonso} 
\affiliation{Instituto de Ciencia de Materiales de Madrid (ICMM), CSIC, Cantoblanco, 28049 Madrid, Spain.}

\date{\today}

\begin{abstract}

We have synthesized Sr$_6$Co$_5$O$_{15}$, a quasi-one-dimensional oxide, measured its magnetic properties and calculated its electronic structure by ab initio techniques. We have found strong evidence for its electronic and magnetic behavior not to follow the trend of its structural series. The magnetic coupling inside the CoO$_3$ chains is not purely ferromagnetic, the long-range coupling inside them is very weak. The Co moments are slightly canted due to their large orbital angular momenta being oriented along each particular quantization axis, that is different for each Co$^{4+}$ atom in the structure. Our thermopower calculations are in agreement with the experiment, supporting our model of the magnetic ground state of the compound.

\end{abstract}
\maketitle

\section{Background}
Transition metal oxides have drawn the attention of the scientific community for the last 50 years. Particularly, cobalt oxides are becoming increasingly important because of their interesting properties such as superconductivity,\cite{superconductivity} colossal magnetoresistance \cite{colossal} or phase separation.\cite{phase}

To elaborate models for strongly correlated electron materials, one-dimensional (1D) systems are key since they are the easiest to study because all the interesting phenomena take place along one direction. In this regard, there has been much interest in analyzing the homologous series A$_{n+2}$B$'$B$_n$O$_{3n+3}$\cite{sugiyama,est_1,est_3} (A alkaline or alkaline earth cations, B$'$ and B commonly corresponding to Co cations in a trigonal prismatic and octahedral position, respectively, and n$\in$[1,$\infty$)) where the one-dimensional chain is represented by B$'$B$_n$O$_{3n+3}$. Boulahya \textit{et al.} \cite{est_2}, reported the integer terms of the series which can be stabilized with Co occupying both octahedral and prismatic sites, by varying the nature and the proportion of alkaline-earth cations. 

In particular, both the end members of the series (Ca$_{3}$Co$_{2}$O$_{6}$ (n=1) and BaCoO$_{3}$ (n=$\infty$)) have focused much attention for the past years. The n=1 compound (Ca$_{3}$Co$_{2}$O$_{6}$) has been analyzed in several previous works.\cite{sugiyama,cacoomaignan,cacoo_aasland,cacoo_aasland_2,cacoo_kageyama,cacoo_fontcuberta,cacoo_khomskii,cacoo_goodenough} The valence state of Co ions in this case is assigned to be 3+, with low spin state for Co ions in the CoO$_6$ octahedron (S=0) and high spin state within the trigonal prism (S=2). This compound shows a paramagnetic behavior at high temperature. In-chain ferromagnetic (FM) interactions arise below 80~K, reaching 1D FM order at 30~K. Below this temperature, interchain two-dimensional (2D) antiferromagnetic (AFM) interactions appear, evolving into a ferrimagnetic order below 24~K (FM order within the chains that are partly antiferromagnetically coupled). According to Wu \textit{et al.},\cite{cacoo_khomskii} the FM intrachain interactions  obey an Ising type model due to the strong spin-orbit coupling effects on the Co ions within a trigonal prismatic environment. However, Cheng \textit{et al.}\cite{cacoo_goodenough} affirm that the FM intrachain-AFM interchain competition invalidates the application of an Ising model. This material has attracted much attention due to the magnetization plateaus observed in the magnetization versus field curves.\cite{cacoomaignan}

The n=$\infty$ member (BaCoO$_{3}$) has also drawn considerable interest.\cite{bacoo_est,bacoo_yamaura,bacoo_struct,bacoo_abinit,bacoo_clusters,sugiyama,bacoo_ps,bacoo_2DAF,bacoo_vpardo} It crystallizes in a 2H hexagonal pseudo-perovskite structure, in which there are just face-sharing CoO$_6$ octahedra forming the 1D CoO$_3$ chain. In this case, the valence state of Co atoms is 4+. \textit{Ab initio} calculations predicted a FM ground state along the chains.\cite{bacoo_abinit} The c-axis is assigned to be an easy direction for the magnetization, leading to a large value of the orbital angular momentum. Also a large, Ising-type, magnetocrystalline anisotropy is estimated.\cite{bacoo_vpardo}

The introduction of prisms in the structure of 2H-BaCoO$_3$ is accompanied by a decreasing of the c$_{2H}$ parameters. Keeping B= Co, to increase the P/O ratio, (P= prism, O= octahedra) control of both temperature and annealing time besides the adequate selection of A cation is required to prepare these materials. The distance between A cations is one of the main factors governing the structural type which can be stabilized in the A$_{n+2}$B$'$B$_n$O$_{3n+3}$ series.

For this reason, there are fewer works on the magnetic or electronic structure properties for the compounds with 2$\leq$n$\leq$$\infty$. The work by Sugiyama \textit{et al.} \cite{sugiyama} studied the electronic structure and magnetic properties of the members n= 1,~2,~3,~5, and $\infty$. The study reported an n-dependence of the charge and spin distribution of the Co chains. They proposed a charge distribution within the chains based on Co$^{4+}$ cations (located in octahedra) and Co$^{3+}$ ones (located in both trigonal prisms and octahedra). The existence of a magnetic transition was shown for all the compounds. Above the temperature atributed to this transition (T$_C^{on}$), a relatively strong 1D FM order appears. They suggested that, for the compounds with n=1,~2,~3, and 5, T$_C^{on}$  is induced by an interchain 2D AFM interaction. Another magnetic study reported by Sugiyama \textit{et al.}\cite{sugiyama2} confirmed the role of this 2D AFM interaction in the series. The structural work carried out by Harrison \textit{et al.}\cite{harrison} for the Sr$_6$Co$_5$O$_{15}$ phase (n=4) reported that this compound is a 2H-hexagonal perovskite related oxide, isostructural with Ba$_6$Ni$_5$O$_{15}$, phase described by Camp\'a \textit{et al}.\cite{campa} Structure, magnetic properties and electronic structure of single crystals of the oxygen-deficient compound Sr$_6$Co$_5$O{$_{14.7}$} have been studied by Sun \textit{et al.}\cite{sunand} They showed this compound to have unique polyhedral chains, consisting of a random composite of octahedra+trigonal prisms and octahedra+intermediate polyhedra. The magnetic properties of the compound can be understood according to that structural picture being the Co$^{4+}$ ions located in the octahedra and the Co$^{2+}$ ones in either the trigonal prisms or the intermediate polyhedra.
Whangbo \textit{et al.} \cite{srcoo_magn} proposed an interpretation of the electronic structure of Sr$_6$Co$_5$O{$_1$$_5$} using a H\"uckel tight binding calculation.\cite{huckel} According to their model, the polyhedral chains are composed by Co$^{4+}$ ions in the octahedral sites and Co$^{2+}$ ions in the trigonal prismatic ones. 
The electrical resistivity and Seebeck coefficient dependence with the temperature of Sr$_6$Co$_5$O{$_1$$_5$} have been measured in some previous works\cite{iwasaki,takami_2} and also for the closely related compound (Sr$_{0.75}$Ba$_{0.25}$)$_6$Co$_5$O{$_1$$_5$}.\cite{takami}

The purpose of this paper is to analyze the electronic structure and special magnetic properties of the n=4 member of the series, Sr$_6$Co$_5$O{$_1$$_5$}. We have synthesized polycrystalline samples of the compound, characterized and analyzed its magnetic properties experimentally. Moreover, we have studied its electronic structure by ab initio methods, analyzing the plausible magnetic configurations and obtaining the magnetic ground state of the system. Also, we have calculated the thermopower using the standard Boltzmann transport theory based on the electronic structure obtained by first principles.

\section{Experimental and computational details}

The SrCoO$_{3-\delta}$  , ``H'' polymorph, Ba$_6$Ni$_5$O$_{15}$-like, was obtained in polycrystalline form by a citrate technique. Stoichiometric amounts of analytical grade Sr(NO$_3$)$_2$ and Co(NO$_3$)$_2.6$H$_2$O were dissolved in citric acid. The solution was slowly evaporated, leading to an organic resin which was dried at 140 $^\circ$C and slowly decomposed at 600 $^\circ$C for 12 h. The sample was then heated at 900 $^\circ$C in air. The hexagonal phase was obtained by slowly cooling in the furnace. The reaction product was characterized by X-ray diffraction (XRD) for phase identification and to asses phase purity. The characterization was performed using a Bruker-axs D8 diffractometer (40 kV, 30 mA) in Bragg-Brentano reflection geometry with Cu K$\alpha$ radiation.

Neutron powder diffraction (NPD) diagrams were collected at the Institut Laue-Langevin, Grenoble (France). The diffraction patterns were acquired at the high-resolution D2B diffractometer with $\lambda$ = 1.594 \AA, at 295 K and at 5 K in the angular range 10$^\circ$ $<$ 2$\theta$ $<$ 156$^\circ$ with a 0.05$^\circ$ step. NPD diffraction patterns were analyzed by the Rietveld method,\cite{rietveld} using the FULLPROF refinement program.\cite{carvajal} A pseudo-Voigt function was chosen to generate the line shape of the diffraction peaks. The coherent scattering lengths for Sr, Co, and O were: 7.020, 2.490, and 5.803 fm respectively. The following parameters were refined in the final run: scale factor, background coefficients, zero-point error, pseudo-Voigt corrected for asymmetry parameters, positional coordinates, and isotropic thermal factors.

The magnetization (M) between 5 and 320 K was measured in a superconducting quantum interference device magnetometer (Quantum Design) under a dc magnetic field H= 100 Oe. Data were taken upon heating in both zero-field-cooling (ZFC) and field-cooling (FC) regimes.


The electronic structure calculations were performed with the WIEN2k code,\cite{wien2k} based on density functional theory (DFT) utilizing the augmented plane wave plus local orbitals method (APW+lo). For the calculations of the transport
properties we used the BoltzTraP code \cite{boltztrap}, that
takes the energy bands obtained using the WIEN2k
software.

For this moderately correlated transition metal oxide, we used the LDA+U \cite{sic} approach including self-interaction corrections in the so-called ``fully localized limit'' with U= 4.8 eV and J= 0.7 eV. This method has proven reliable for transition metal oxides, since it improves over the generalized gradient approximation (GGA) or local density approximation (LDA) in the study of systems containing correlated electrons by introducing the on-site Coulomb repulsion U. Results presented here are consistent for values of U in the interval from 4 to 8 eV, in a reasonable range compared to other similar cobaltates,\cite{cacoo_khomskii,bacoo_vpardo} to describe correctly the insulating behavior of the material and the localized nature of its electronic structure.

The calculations were fully converged
with respect to the k-mesh and R$_{mt}$K$_{max}$.
Values used for the k-mesh were 6$\times$6$\times$6 sampling of the full Brillouin zone for electronic structure calculations, and
21$\times$21$\times$21 for the transport properties.~R$_{mt}$K$_{max}$= 6.0
is chosen for all the calculations. Selected muffin tin radii were
the following: 1.82 a.u. for Co, 2.28 a.u. for Sr, and 1.61 a.u. for O.
Based on scalar relativistic basis functions, spin orbit coupling (SOC) effects  were included
in a second-variational procedure.\cite{singh}

\section{Results}

\subsection{Structure}

The crystal structure (see Table \ref{positions}) of this phase  was refined by Rietveld analysis of the NPD data and published by us \cite{cristina}  in the R32 space group (no.~155), Z = 3, at room temperature and at 5 K,  starting from the model defined by Harrison \textit{et al.}\cite{harrison} It contains two strontium, three cobalt, and three oxygen atoms in the asymmetric unit. The refinement of the occupancy factor for the Co atoms leads to a significant reduction of its contents for Co1, whereas Co2 and Co3 remained stoichiometric. The oxygen occupancy was slightly deficient for O2 and fully stoichiometic for O1 and O3. The refined crystallographic formula was SrCo$_{0.78(1)}$O$_{2.48(2)}$. According to this formula, the average oxidation state for Co is 3.79(1)+. The presence of Co$_3$O$_4$, segregated from the main phase during the synthesis process, was quantified as 2\%. 

\begin{figure}
\includegraphics[width=\columnwidth,draft=false]{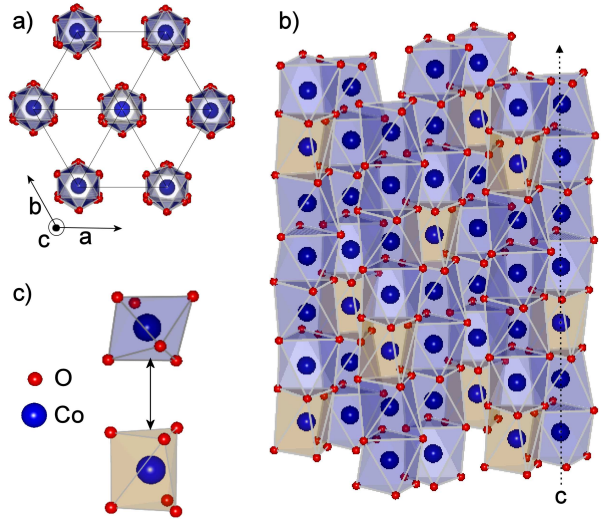}
\caption{(Color online) a) Top view of the structure of Sr$_6$Co$_5$O$_{15}$ showing the hexagonal symmetry of the ab plane, and the 6-fold coordination of the Co atoms by O atoms. The in-plane distance between Co chains is significantly larger than the Co-Co in-chain distance, leading to the structural quasi-one-dimensionality. Sr atoms are not shown for simplicity. b) Schematic picture of the structure of the CoO$_3$ chains in Sr$_6$Co$_5$O$_{15}$ showing the polyhedral environment of the different Co cations that occur along them. In the unit cell, formed by 5 Co atoms, 4 of them are situated in a distorted octahedron (blue color) and one is in a trigonal prismatic environment (yellow color). c) Detail of the face-sharing arrangement along the chains.}\label{struct}
\end{figure}

Both Sr atoms are 8-fold coordinated by O atoms ($\langle$Sr1-O$\rangle$ = 2.592(2) \AA, $\langle$Sr2-O$\rangle$ = 2.751(2) \AA) in irregular coordination. The atoms Sr1 and Sr2 are disposed in columns parallel to the c direction. Two cobalt atoms, Co2 and Co3 (site 6c) are octahedrally coordinated by oxygen atoms ($\langle$Co2-O$\rangle$ = 1.896(6) \AA,  $\langle$Co3-O$\rangle$ = 1.905(6) \AA) and the third, Co1 (site 3b) occupies a distorted trigonal prism  ($\langle$Co1-O$\rangle$ = 1.922(2) \AA~and O-Co1-O = 76.8(4)$^\circ$). The crystal structure consists of isolated, infinite chains of face-sharing CoO$_6$ polyhedra running along the c direction forming a trigonal lattice in the ab plane as can be seen in Fig.~\ref{struct}a). The repeat unit for Co-O species consists of four distorted octahedra sharing faces intermingled with prismatically-coordinated Co1 atoms (see Fig.~\ref{struct}b),c)). The inter-octahedral cobalt-cobalt distances are:  Co2-Co3 = 2.39(3) \AA, Co2-Co2 = 2.54(3) \AA. The face-sharing prismatic/octahedra Co1-Co3 distance is 2.53(2) \AA. These results \cite{cristina} are comparable with those obtained by Harrison \textit{et al.}\cite{harrison} for the Sr$_6$Co$_5$O$_{15}$ phase, although in our case we are in the presence of a more severely Co deficient compound, with a slightly smaller cell volume of 968.3 \AA$^3$ (969.6 \AA$^3$ for Sr$_6$Co$_5$O$_{15}$ \cite{harrison}) as corresponding to a higher average oxidation state for Co cations. The in-plane distance between Co chains is around 5.62 \AA~in the ``ideal'' structure and is flanked, in the refined structure, by 5.54 and 5.75 \AA~(measured between different pairs of Co atoms belonging to two neighboring chains).

\begin{table}[h!]
\caption{Atomic positional parameters for Sr$_6$Co$_5$O$_{15}$ ``H'' phase after Rietveld refinement of NPD data at 5K. The space group of our compound is R32 (no. 155) and the lattice parameters are a=~b=~9.4740 \AA ~and c=~12.3606 \AA.}\label{positions}
\begin{center}
\begin{tabular}{c c c c c}
\hline
\hline

 &   Crystallographic& \\ 
Atom &   position  & Coordinates \\
\hline
Sr1 &   9e & (0.6437,0.0000,0.5000) \\
Sr2 &   9d & (0.3210,0.0000,0.0000) \\
Co1 &  3b &  (0.0000,0.0000,0.5000) \\
Co2 &   6c & (0.0000,0.0000,0.1030) \\
Co3 &   6c & (0.0000,0.0000,0.2960) \\
O1 &   9d & (0.8436,0.0000,0.0000) \\
O2 &   18f &  (0.4946,0.6728,0.4785) \\
O3 &   18f & (0.8436,-0.0240,0.6088) \\

\hline
\end{tabular}
\end{center} 
\end{table}

\subsection{An ionic model}

Being the compound a correlated insulating oxide,\cite{iwasaki} an image based on an ionic point charge model (PCM) can give us a crude estimate of the possible electronic configuration of the material. We will use such a model to describe the charge distribution of the cations along the Co chain. We will consider possible ionic configurations for the different Co ions along the chain and calculate their total energy, just based on the electrostatic repulsion, simplifying to take into account only the first neighbor contribution, and neglecting other energetic terms. 


Taking the usual valencies for Sr and O, the average valence for Co in the ideal stoichiometric compound Sr$_6$Co$_5$O$_{15}$ is +3.6. Following the PCM we have just described, the valencies of the Co ions can be distributed in two isoenergetic ways in order to minimize the Coulomb repulsion: i) 4 Co$^{4+}$ (d$^{5}$)+ 1 Co$^{2+}$ (d$^7$) ; ii) 3 Co$^{4+}$ (d$^5$)+ 2 Co$^{3+}$(d$^6$). Various ionic arrangements have been considered in the literature. According to the structure determined by Harrison \textit{et al.},\cite{harrison} Sun \textit{et al.}\cite{sunand} proposed 4 Co$^{4+}$+ 1 Co$^{2+}$ is the most suitable model because the polyhedral chain with 4 octahedral sites and one trigonal prismatic allows Co$^{4+}$ and Co$^{2+}$ to be located in different sites. 
Whangbo \textit{et al.}\cite{srcoo_magn} also suggested the 4 Co$^{4+}$ (d$^{5}$)+ 1 Co$^{2+}$ (d$^7$) model using a H\"uckel tight binding calculation. This would be what we called solution i). Instead, Sugiyama \textit{et al.} \cite{sugiyama} proposed the existence of at least one non-magnetic atom in the chain for all the members of the series A$_{n+2}$Co$_{n+1}$O$_{3n+3}$: as n increases from 1 up to infinity, the Co valence increases from +3 and approaches +4 (e.g. for n=1, the charge distribution in the unit cell is 2 Co$^{3+}$; for n=2, 2 Co$^{3+}$+ 1 Co$^{4+}$ and in our case for n=4, 2 Co$^{3+}$+ 3  Co$^{4+}$). 
This, we called solution ii). In addition, in Ref.~[\onlinecite{sugiyama}] is suggested that the spin distribution for the 2 Co$^{3+}$ ions in the chain is: a high spin state (HSS) with S=2 for the Co within a trigonal prismatic environment and a low spin state (LSS) with S=0 for the octahedral one. Meanwhile, the Co$^{4+}$ ions are all in a LSS with S=1/2 and located in the remaining octahedra. 

Both solutions are equivalent energetically from an oversimplified ionic picture, but can be easily distinguished because solution ii) could lead to two non-magnetic Co atoms, whereas solution i) will have all the atoms being magnetic.
Our ab initio calculations confirm that the ground state electronic structure can be well described by an ionic model with 3 Co$^{4+}$ and 2 Co$^{3+}$ cations (solution ii) of our PCM). Below, we will give further details of the electronic structure beyond this simple ionic model.

\subsection{Electronic structure calculations}

\begin{figure}
\includegraphics[width=0.90\columnwidth,draft=false]{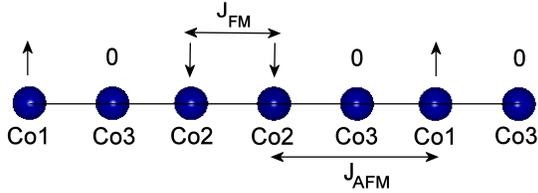}
\caption{(Color online) Magnetic couplings in the unit cell. We show the FM coupling between neighbor Co2 atoms and the AFM one between Co2 and Co1 mediated by a non-magnetic Co3.}\label{acoplos}
\end{figure}

Table~\ref{momentos_tabla} shows the magnetic moments of each Co cation in the structure obtained for U= 4.8 eV. They are consistent with the ionic distribution that would predict two non-magnetic Co$^{3+}$:~d$^6$ cations to occur along the chain. Co1 is a Co$^{4+}$:~d$^5$ cation with a magnetic moment of 1 $\mu_B$. Co2 is also a Co$^{4+}$:~d$^5$ cation and Co3 is close to a Co$^{3+}$:~d$^6$ configuration. The details can be understood looking at the magnetic interactions in the chain (see Fig.~\ref{acoplos}). The magnetic coupling between Co1 and Co2 is mediated by a Co3 (non magnetic). The overlap between Co3 and Co2 d-orbitals motivates a charge transfer that explains the magnetic moments obtained for them (lowered from 1 for Co2 (Co$^{4+}$:~d$^5$) and risen from 0 for Co3 (Co$^{3+}$:~d$^6$)). The Co$^{4+}$ cations are in a LSS (S=1/2) in agreement with Ref.~[\onlinecite{sugiyama}]. However, for our compound, both the Co$^{3+}$ atoms are non-magnetic (LSS) and located in octahedra.

\begin{table}[h!]
\caption{Projection of the spin magnetic moments of Co atoms in the Sr$_6$Co$_5$O$_{15}$ ground state.}\label{momentos_tabla}
\begin{center}
\begin{tabular}{c c c c c}
\hline
\hline
Atom  & & Magnetic Moment & \\
\hline
Co1 & & 1.0$~\mu_B$ \\
Co2 & & -0.8$~\mu_B$ \\
Co3 & & -0.2$~\mu_B$ \\

\hline
\end{tabular}
\end{center} 
\end{table}

\begin{figure}
\includegraphics[width=\columnwidth,draft=false]{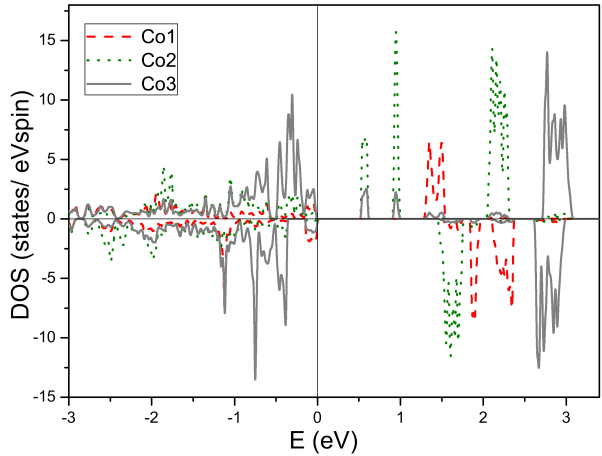}
\caption{(Color online) Partial spin-polarized DOS of Co1, Co2, and Co3 atoms. Fermi energy is represented by the solid vertical line at zero. Co1 and Co2 are close to a d$^{5}$ electronic structure. Co3 is closer to a d$^{6}$ configuration with a little unoccupied t$_{2g}$ character.}\label{dos}
\end{figure}

A more realistic description of the electronic structure of the material is given by the partial density of states (DOS) plots of the various Co atoms in the structure (see Fig.~\ref{dos}). The material is an insulator, with a d-d gap of about 0.5 eV, for this particular value of U (4.8 eV). For Co1, a Co$^{4+}$:~d$^5$ cation in a trigonal prismatic environment, we can see a full d$_{z^2}$ level (being z the Co-chain axis) fully occupied, a hole in an xy-plane orbital (x$^2$-y$^2$, xy) at about 2 eV above the Fermi level and the higher-lying d$_{xz}$, d$_{yz}$ which remains unoccupied at higher energy, spin split by about 1 eV. The bands coming from Co2 (Co$^{4+}$:~d$^5$), in an octahedral environment, present a spin splitting of about 1 eV of the e$_g$ bands, located at 2 eV above the Fermi level. In this case, the magnetic moment points along the minority spin direction (the hole in the t$_{2g}$ multiplet is in the majority spin channel). The unoccupied t$_{2g}$ band of Co2 presents a double  peak structure at about 1 eV above the Fermi level. We can observe an approximate d$^6$ DOS for Co3. Due to the hybridization between Co3 and Co2 d-orbitals, a double peak structure arises, showing a density of unoccupied t$_{2g}$ states for Co3 much smaller than for Co2 at about 1 eV above the Fermi level, consistent with the small magnetic moment of Co3.

Two features can be observed in these plots: i) the strongly localized nature of the electrons, with very narrow bands, less than 0.5 eV wide. The band widths are always smaller than the typical energies involved: both the Hund's rule (1 eV for the e$_g$ bands of the Co$^{4+}$:~d$^5$ cations) and crystal field splitting, ii) the different crystal field environments of the Co cations (octahedral for Co2 and Co3, trigonal prismatic for Co1), that lead to well different splittings with the e$_g$ states of Co3 being highest in energy, at $\approx$ 3~eV above the Fermi level for this U value chosen of 4.8 eV.

From these results, we can roughly sketch the electronic structure of Sr$_6$Co$_5$O$_{15}$: its unit cell is formed by three magnetic Co$^{4+}$:~d$^5$ cations (one Co1 in a trigonal prismatic environment and two Co2 in an octahedral environment) and two non-magnetic Co$^{3+}$:~d$^6$ atoms (Co3) in an octahedral environment.

\subsection{Magnetic properties}

We have performed LDA+U calculations for several values of the on-site Coulomb repulsion term. Since we are dealing with an insulating d$^5$/d$^6$ system, a value of U between 4 and 8 eV is reasonable to describe it correctly.\cite{ylvisaker} The magnetic ground state solution we will describe below (magnetic moments of Co atoms in the ground state are written in Table~\ref{momentos_tabla}) is the most stable for the said range of values of the on-site Coulomb repulsion.  

Starting from the electronic structure described above, we can understand the magnetic couplings in the unit cell. Two couplings can be considered (see Fig.~\ref{acoplos}): one ferromagnetic (J$_{FM}$) direct exhange  between nearest neighbor magnetic Co2:~d$^5$ cations and another one antiferromagnetic (J$_{AFM}$) between Co1:~d$^5$ and Co2:~d$^5$ mediated by a non-magnetic cation Co3:~d$^6$, that acts in a similar way to O anions in the oxygen-mediated superexchange in perovskites. Both these couplings can be understood in terms of the Goodenough-Kanamori-Anderson rules.\cite{goodenough_book} We can use our total energy calculations to describe and quantify the magnetic interactions in the unit cell (schematically depicted in Fig.~\ref{acoplos}). In order to give an estimate of the couplings along the chain, we can fit the total energies resulting from various possible collinear magnetic configurations to a Heisenberg model, in the form $H= \frac{1}{2}\sum^{}_{i,j}J_{ij}S_iS_j$. Calculations reveal the following values for the coupling constants J$_{FM}\approx220K$  and  J$_{AFM}\approx6K$ (of opposite sign). As expected, the FM coupling is stronger than the AFM one, that occurs between second-neighbor cations mediated by a non-magnetic ion. The small AFM coupling is consistent with the fact that no long-range magnetic order along the Co-chains is observed experimentally above 4~K. Because of this peculiar magnetic arrangement, no 1D FM order (along the Co-chains) is observed, but only a short-ranged FM coupling between Co2 cations survives at high temperature. The total ordered moment is 1 $\mu_B$ per unit cell.

Figure 4 shows the magnetic susceptibility as a function of the temperature at 100 Oe. At 32 K, a kink can be noticed in the ZFC curve, which can be ascribed to the N\'eel temperature (T$_N$). Also, this is consistent with the significant increase of $\chi$ below this temperature observed in the FC curve.  This magnetic transition could be related to an interchain 2D AFM interaction in the triangular lattice of the ab plane as in the other members of the series.\cite{sugiyama} Confirming this picture, a large negative Curie-Weiss temperature ($\theta= -109$ K) is obtained as in the n=2 and n=3 compounds.\cite{sugiyama} The T$_N$ value given by Sugiyama \textit{et al.}\cite{sugiyama2} for Sr$_6$Co$_5$O$_{15}$ is about 70 K, in agreement with the general magnetic behavior of the series. According to them, this temperature is due to the transition from a 1D FM order along the chains  to a 2D AFM ordered state. In addition, no clear anomalies in the susceptibility vs. temperature curve for the members with n=1, 2, 3, and 5 are observed in Ref.~[\onlinecite{sugiyama}]. In our case, as we have seen above, there is no 1D FM order along the chains. Also, reasoning in terms of the hexagonal planes, non-magnetic planes formed by Co3 atoms intercalate between the magnetic ones. All this contributes to the lowering of T$_N$ with respect to the rest of the series.

\begin{figure}
\includegraphics[width=8cm,draft=false]{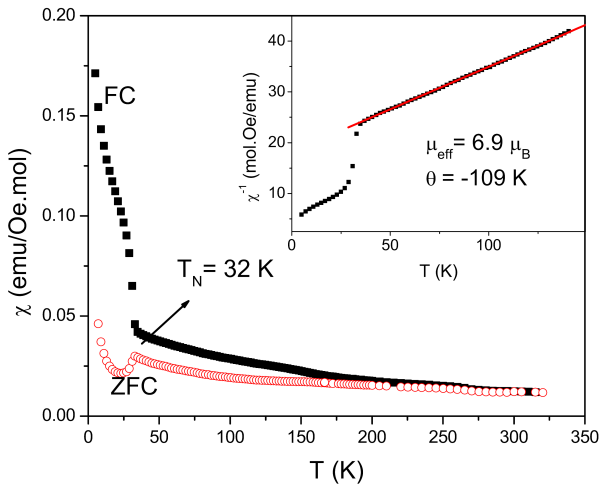}
\caption{(Color online) Magnetic susceptibility vs. temperature measured at H=100 Oe under ZFC and FC conditions. The N\'eel temperature that can be identified from the curves is about 32 K. The inset shows the variation of FC inverse susceptibility at low temperatures. $\theta$ represents the Curie-Weiss temperature.}\label{pablo}
\end{figure}

From our calculations, we can also obtain the value of the effective magnetic moment per formula unit  and compare it with our experimental findings. Hence, taking the usual expressions for the square effective paramagnetic moment of each magnetic Co, $\mu_{eff}^2= [g_l^2l(l+1)+g_s^2s(s+1)]~\mu_B^2$ and considering l= 0 or 1, $\mu\in$[5.2, 6.7]~$\mu_B$ for the whole unit cell formed by 5 Co cations. Depending on the value of the orbital angular momentum (see below the details of our calculations), that would be the range of possible values for $\mu_{eff}$. This is consistent with the 5.6 $\mu_B$ value obtained by Sun el al,\cite{sunand} (if the orbital angular momenta were negligible), and it agrees with our experimental magnetic moment of 6.9 $\mu_B$ per formula (if the orbital angular momenta were aligned with the magnetization). This value was obtained from the linear fitting of $\chi^{-1}$ vs T curve in the paramagnetic region (see inset of Fig. 4).

\subsection{A peculiar quasi-one-dimensional oxide}

The magnetic properties of the material can be placed into the context of the other members of the same structural series.\cite{sugiyama} Keeping in mind the T$_N$ dependence with the in-plane distance for the quasi-one-dimensional cobaltates shown in Ref.~[\onlinecite{sugiyama2}] we can see the T$_N$ value for Sr$_6$Co$_5$O$_{15}$ of about 70 K, sensibly higher than the one obtained experimentally by us (see Fig.~\ref{pablo}) and others.\cite{sunand}

Both the end members of the series have in common an Ising-type behavior,\cite{cacoo_khomskii,bacoo_vpardo} with moments aligned along the chain direction with large values of the magnetocrystalline anisotropy. This quasi-one-dimensionality is somehow not observed in Sr$_6$Co$_5$O$_{15}$. We have studied the system by introducing SOC in the calculations with the magnetization lying along different crystallographic directions. However, none of these directions can align the orbital and magnetic moments of all the Co atoms at the same time (see Table~\ref{momentos_sot}). Large values of the orbital angular momenta are obtained for the Co atoms in the structure when the magnetization is set along different directions: the preferred direction for orienting their moments is different for each magnetic Co (Co1 and Co2) in the unit cell. For them, the orbital angular momentum is parallel to the spin moment, as a result of the Hund's third rule. For Co1, the degenerate x$^2$-y$^2$ and xy levels (where the hole resides) can form a linear combination of eigenstates with l$_z$=2,\cite{cacoo_khomskii} so the Co$^{4+}$:d$^5$ cation in a prismatic environment is susceptible of developing a ground state with a large value of the orbital angular momentum. On the other hand, Co2 (Co$^{4+}$:~d$^5$) can develop an l$_z$= 1 eigenstate since the t$_{2g}$ multiplet acts as an effective l= 1 multiplet.\cite{epr,stevens,enough,lacroix,eschrig} The ground-state quantization axis will be related to the local environment which is rotated for the two Co2 atoms in the unit cell. This explains why we have different l$_z$ values even for the two equivalent Co2 atoms in the structure. Consequently, this spin-system cannot be described as an Ising-type one due to the canting of the moments of the various magnetic ions with respect to each other. Such canting can be understood according to the different Co environments and as a local orientation of the moments along its particular symmetry axis. The values are summarized in Table \ref{momentos_sot}, and help understand the measured $\mu_{eff}$ value.

Another difference with the other compounds in the series is that, in this case, we do not have  a strong in-chain FM coupling for all the Co atoms in the unit cell. Thus, the magnetic properties of Sr$_6$Co$_5$O$_{15}$ cannot be understood as FM spin chains because of the AFM coupling that (though weak) occur within them and the two non-magnetic Co atoms per unit cell.

Clearly, the magnetic properties of this compound differ from the other members of the series making it less quasi-one-dimensional.

\begin{table}
\caption{Projection of the orbital angular momenta of Co atoms along the magnetization axis for different directions of the magnetization (in $\mu_{B}$ units).} \label{momentos_sot}
\begin{ruledtabular}  
	\begin{tabular}{lcccccccc}
		\multicolumn{9}{c}{ Atom \hspace{1cm} l$_{z}$ for various magnetization directions} \\ \cline{1-2} \cline{3-9} 
 &  & (111) &(101)&(110)& (011) &(100)& (010)& (001)\\
\hline
Co1 & & 0.99 & 0.75 & 0.66 & 0.72 & 0.41 & 0.36 & 0.44 \\
Co2(a) & & -0.05 & -0.15 & -0.17 & -0.19 & -0.22 & -0.28 & -0.31 \\
Co2(b) & & -0.05 & -0.20 & -0.18 & -0.17 & -0.28 & -0.29 & -0.27\\ 

\end{tabular}  
\end{ruledtabular}  
\end{table}

\subsection{Transport properties.}

\begin{figure}
\includegraphics[width=\columnwidth,draft=false]{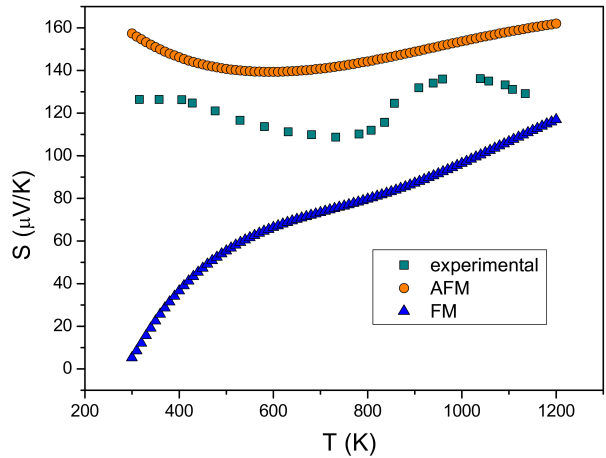}
\caption{(Color online) Experimental \cite{iwasaki} and calculated temperature dependence of the thermopower. The results for both the FM and AFM spin-configurations are plotted.}\label{thermopower}
\end{figure}

We have calculated the thermoelectric power dependence with the temperature to further analyze the system properties and the magnetic ground state. This has been done taking our band structure calculations within a semiclassical approach based on the Boltzmann transport theory through the BoltzTraP code.\cite{boltztrap} A dense grid of 10000 k points in the full Brillouin zone has been used to obtain convergence. Taking the conductivity ($\sigma$) and Seebeck coefficient ($S$) calculated for both spin channels, the total thermopower has been obtained according to the two-current model expression:\cite{singh_nacoo}

\begin{equation}
S=\frac{\sigma'(\uparrow)S(\uparrow)+\sigma'(\downarrow)S(\downarrow)}{\sigma'(\uparrow)+\sigma'(\downarrow)}
\end{equation}
where $\sigma'$=~$\sigma/\tau$, within the constant scattering time ($\tau$) approximation. For the sake of comparison, we present the calculations obtained for the ground state AFM spin configuration shown in Fig.~\ref{acoplos} and also for a FM solution (which is higher in energy according to our calculations). We present the data calculated for both spin configurations at U= 4.8 eV, together with the experimental values of the Seebeck coefficient taken from Ref.~[\onlinecite{iwasaki}] (see Fig.~\ref{thermopower}). The results for the AFM solution fit the experimental values nicely, both in order of magnitude of the thermopower and also on the observed non-activated evolution with the temperature. However, the FM one differs clearly from the experiment, giving further evidence of the validity of our description of the electronic structure of the compound.

\section{Summary}

We have synthesized Sr$_6$Co$_5$O$_{15}$, measured its magnetic properties and performed ab initio calculations. The material is structurally quasi-one-dimensional and can be identified as a member of the structural series A$_{n+2}$B$'$B$_n$O$_{3n+3}$ of hexagonal quasi-one-dimensional Co oxides. It is an insulating antiferromagnet with T$_N$= 32 K, which does not correspond to the magnetic properties expected for a member of the series with that value of the in-plane Co-Co distance. This is due to its peculiar electronic structure properties: i) the in-chain couplings are not purely FM. AFM couplings (though weak) occur within the chains. ii) This, together with the existence of two non-magnetic Co atoms in the unit cell, reduces the inter-chain magnetic couplings, responsible of the higher T$_N$ in the other hexagonal quasi-one-dimensional cobaltates. iii) The preferred orientation of the orbital angular momenta is non-collinear, in contrast to the strong Ising-type behavior found in Ca$_3$Co$_2$O$_6$ and BaCoO$_3$. In addition, transport properties calculations support our understanding of its electronic structure and magnetic properties.

This anomalous behavior makes Sr$_6$Co$_5$O$_{15}$ less quasi-one-dimensional than expected due to its structure and helps understand the intricate structure-property relations in strongly correlated electron systems.

\section{Acknowledgments}

The authors thank the CESGA (Centro de Supercomputaci\'on de Galicia) for the computing facilities and the Ministerio de Educaci\'{o}n y Ciencia (MEC) for the financial support through the project MAT2009-08165. Authors also thank the Ministerio de Ciencia e Innovaci\'on (MICINN) for the project MAT2007-60536 and the Xunta de Galicia for the project INCITE08PXIB236053PR. A.~S. Botana thanks MEC for a FPU grant.

\end{document}